\documentclass{article}
\usepackage{spconf,amsmath,graphicx}
\usepackage{color}

\title{Estimating Indoor Scene Depth Maps from Ultrasonic Echoes}
\name{Junpei Honma$^\dag$~~~~~Akisato Kimura$^\ddag$~~~~~Go Irie$^\dag$}
\address{\dag Tokyo University of Science~~~~~\ddag NTT Corporation}

\begin{document}
%
\maketitle
\begin{abstract}
Measuring 3D geometric structures of indoor scenes requires dedicated depth sensors, which are not always available.
Echo-based depth estimation has recently been studied as a promising alternative solution.
All previous studies have assumed the use of echoes in the audible range. However, one major problem is that audible echoes cannot be used in quiet spaces or other situations where producing audible sounds is prohibited.
In this paper, we consider echo-based depth estimation using inaudible ultrasonic echoes.
While ultrasonic waves provide high measurement accuracy in theory, the actual depth estimation accuracy when ultrasonic echoes are used has remained unclear, due to its disadvantage of being sensitive to noise and susceptible to attenuation.
We first investigate the depth estimation accuracy when the frequency of the sound source is restricted to the high-frequency band, and found that the accuracy decreased when the frequency was limited to ultrasonic ranges.
Based on this observation, we propose a novel deep learning method to improve the accuracy of ultrasonic echo-based depth estimation by using audible echoes as auxiliary data only during training.
Experimental results with a public dataset demonstrate that our method improves the estimation accuracy.

\end{abstract}
%
\section{Introduction} \label{sec:intro}
The geometric structure of a scene is essential for a variety of applications, including navigation, path planning for autonomous mobile robots, and spatial layout design for indoor scenes.
Measuring geometric structures requires specialized optical sensors to acquire the depth of a scene, such as infrared light, LiDARs, or specially configured camera devices such as stereo cameras.
However, such measurement devices are not always available as these are often costly and require strict setup conditions for accurate measurements.
While deep monocular depth estimation, which uses deep learning to estimate depth maps of scenes from monocular RGB images captured by ordinary cameras, has also been explored for a decade~\cite{eigen2014depth,eigen2015predicting,laina2016deeper,ma2018sparse}, there are many spaces where cameras cannot be used, such as dark rooms or spaces with privacy protection or legal restrictions.

\begin{figure}[t]
  \centering
  \includegraphics[width=0.9\linewidth]{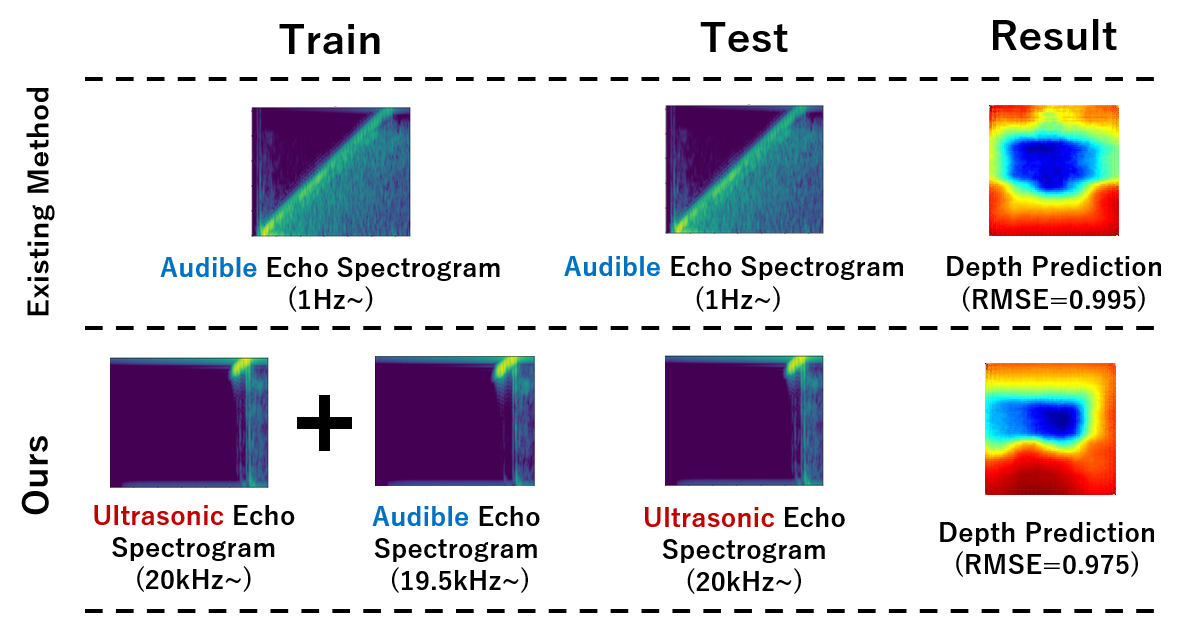}
  \vspace{-12pt}
  \caption{{\bf Overview of Our Idea.} 
  Top: All existing echo-based depth estimation methods use audible echo spectrograms during training and testing, which are not always used depending on the surrounding conditions of the target scene.
  Bottom: In this paper, we aim to mitigate this problem by using inaudible ultrasonic echo spectrograms during testing. 
  Based on the observation that a straightforward approach that restricts the frequency band to the ultrasonic range leads to poor depth estimation accuracy, we propose an approach that uses audible echoes as auxiliary data only during training. We confirm that our method improves the depth estimation accuracy in terms of root mean squared error (RMSE) between the estimated and the ground truth depth maps.}
  \vspace*{-0.45cm}
  \label{fig:teian1}
\end{figure}

In this study, we consider echo-based depth estimation.
Suppose we have a microphone array consisting of multiple microphones at different spatial locations in the scene.
A known sound emitted from a sound source (i.e, loudspeaker) is reflected by surfaces such as walls, windows, furniture, etc., and arrives at each microphone.
The time of arrival to each microphone depends on geometric properties of the surfaces in the scene. That is, the time difference of arrival of the echoes contains information about the geometric structure of the scene.
The problem of recovering the depth map from the echoes is an inverse problem and is difficult to solve analytically, hence is usually solved using deep learning.

Several efforts on echo-based depth estimation have been reported in the literature~\cite{haahr2020batvision,vasudevan2020semantic,gao2020visualechoes,turpin2021sparse,parida2021beyond,irie2022co,zhang2022stereo,wang2023multiModal,zhu2023environment,yan2021rigNet,yun2023dense}.
These primarily focus on exploring effective network architectures for this task.
For example, the use of U-Net~\cite{haahr2020batvision,parida2021beyond}, spatial pyramid pooling~\cite{vasudevan2020semantic}, and bilinear attentions~\cite{irie2022co} have been investigated.
Besides these, multi-modal approaches combined with RGB images have also been discussed~\cite{gao2020visualechoes,parida2021beyond,zhang2022stereo,wang2023multiModal,zhu2023environment,yan2021rigNet,yun2023dense}.
However, one major drawback of the existing methods is that they all assume the use of audible echoes observed using audible sound sources.
Obtaining effective echoes that can stably acquire the 3D structure of an indoor scene requires generating sound loud enough to reverberate throughout the room.
Therefore, the existing methods cannot be used in rooms where generating audible sound is prohibited or where harmful effects on the surrounding environment or human body are concerned.

In this paper, we examine echo-based depth estimation using inaudible ultrasonic sound sources. To the best of our knowledge, this is the first work to explore ultrasonic echo-based depth estimation. 
On one hand, an ultrasonic wave has a short wavelength, which has a theoretical potential to provide high measurement accuracy.
On the other hand, however, a critical drawback is that it is sensitive to noise or interference and tends to attenuate quickly.
Due to this nature of ultrasonic echoes, practical applications of ultrasonic measurements in air have been mainly limited to point measurements within a short distance range (typically $<$ 1m), and the actual accuracy in depth estimation, which requires measurements of two-dimensional surfaces in a longer range (typically $<$ 10m~\cite{eigen2014depth}), has remained unknown.
Therefore, we first conduct preliminary experiments to investigate how the depth estimation accuracy changes when the frequency of the sound source is gradually limited from the audible range to the ultrasonic range.
From the results we found that the estimation accuracy decreased when the frequency range was limited to the ultrasonic band only (discussed later in Sec.~\ref{sec:prelimininary}).

In light of the finding, we propose a novel deep learning method that uses audible echoes as auxiliary data only during training. 
Our method generates synthetic echoes for training by linearly mixing the spectral information of ultrasonic and audible echoes~\cite{zhang2018mixup}, and uses the synthetic echoes as auxiliary data (Fig.~\ref{fig:teian1}). This enables learning of a depth estimation network that is robust to missing audible frequency bands.
Experimental results with Replica~\cite{straub2019replica}, which is one of the most popular public datasets for echo-based depth estimation, demonstrate that our method improves the depth estimation accuracy using ultrasonic echoes.

\section{PRELIMINARY EXPERIMENTS} \label{sec:prelimininary}

We first conduct preliminary experiments to assess the viability of ultrasonic sound sources in the context of depth estimation for indoor scenes. 
Specifically, we evaluate the depth estimation accuracy when the frequency band of the sound source is gradually limited toward the ultrasonic band.

\begin{figure}[t]
    \centering
    \includegraphics[width=0.9\linewidth]{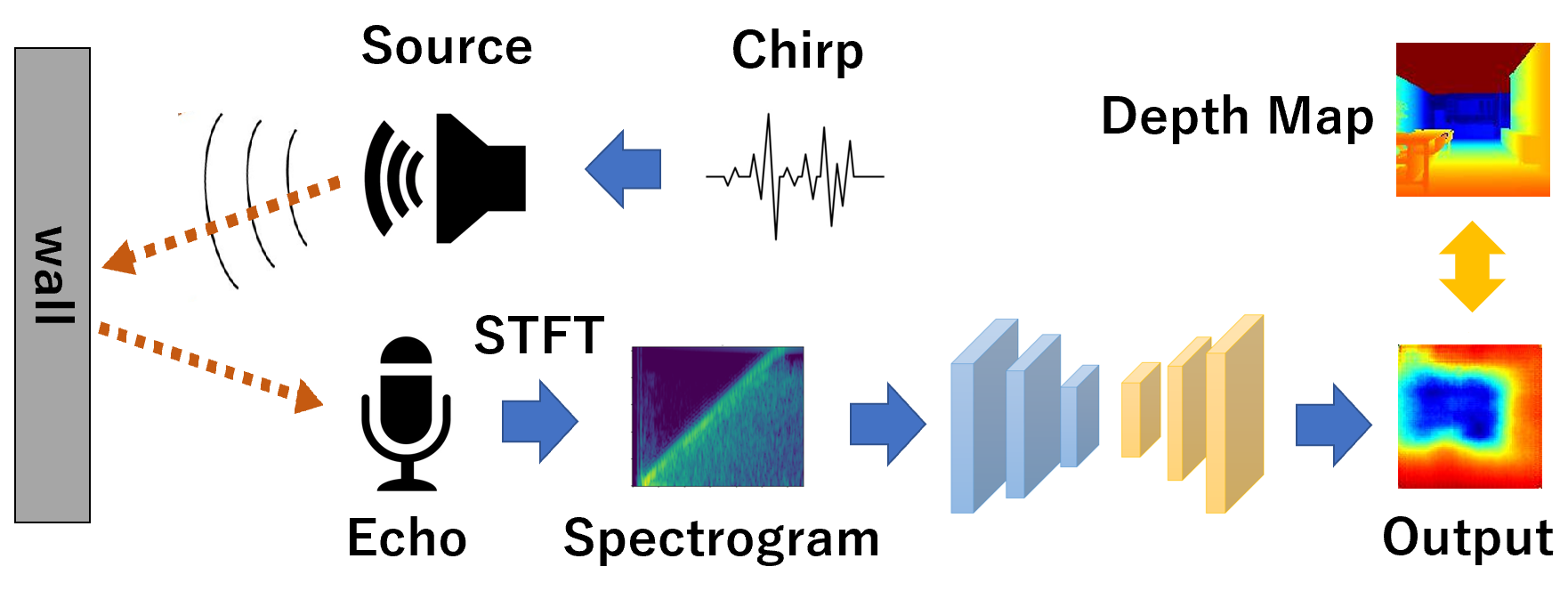}
    \vspace{-15pt}
    \caption{{\bf Echo-based Depth Estimation Framework.} 
    A known chirp signal is emitted to the indoor scene and spectrograms of the multi-channel echoes from the microphone array are extracted.
    The features are fed into a convolutional neural network (CNN) to estimate the depth map of the scene.
    The CNN is trained to minimize the RMSE between the estimated and ground truth depth maps.} 
    \vspace*{-0.3cm}
    \label{fig:nagare}
\end{figure}

\subsection{Dataset} \label{sec:dataset}
We consistently use Replica~\cite{straub2019replica}, one of the two standard public benchmark datasets for evaluating the accuracy of echo-based depth estimation~\cite{parida2021beyond,irie2022co}\footnote{The other dataset, Matterport3D~\cite{Chang2017matte}, does not provide room impulse responses that can reproduce ultrasonic bands so cannot be used for this paper.}.
Replica has data of a total of $18$ indoor scenes covering hotels, apartments, office rooms, etc, and this has been used in research using machine learning for egocentric computer vision, semantic segmentation in 2D and 3D, and performing navigation development. For training and testing, we follow the official training-test split provided by the dataset publisher. The original Replica dataset publishes ground truth depth maps and binaural echoes for various spatial locations and orientations within each scene. However, the echoes provided are limited to audible range up to $16,000$ Hz, so experiments with ultrasonic echoes cannot be conducted.
We therefore synthesize ultrasonic echoes by using the room impulse responses (RIRs) associated with the Replica dataset.
The RIR is the echo observed at a certain location when an impulse signal is emitted at (another) location.
Hence, by convolving the RIR with the input sound emitted at the location of the sound source, the echo observed at a given location can be simulated.
More specifically, let $h(t)$ and $x(t)$ denote the RIR and the sound emitted at the location of the sound source, respectively. The echo $y(t)$ is simulated by the following equation.
\vspace*{-0.2cm}
\begin{equation} \label{eq:rir}
y(t) = \sum^{t}_{k=0} h(t-k)x(k) 
\end{equation}
We use a chirp that varied from $1$ Hz to $22,050$ Hz in $0.05$ seconds  as the sound source and apply a high-pass filter to it to limit the frequency range. In order to account for higher-order reverberation components, we use a sampling frequency of 44,100 Hz and a sufficiently long period of recording time ($0.12$ seconds).
The synthesized echoes are used as input for estimation, and the depth map provided by the original Replica dataset is used as the ground truth output.

\subsection{Depth Estimation Method}
We design our depth estimation framework shown in Fig~\ref{fig:nagare} by following the state-of-the-art echo-based depth estimation method~\cite{parida2021beyond}. Note that several more recent echo-based depth estimation methods have been proposed ~\cite{zhang2022stereo,yan2021rigNet,yun2023dense} but these methods are not applicable to our problem because they assume that special images other than echoes are available as input (stereo images~\cite{zhang2022stereo}, RGB-D images~\cite{yan2021rigNet}, and spherical images~\cite{yun2023dense}).
First, the short-time Fourier transform (STFT) is applied to the binaural echoes generated by the procedure described in Sec.~\ref{sec:dataset} to get the spectrograms of the observed echoes. 
The upper limit of the effective frequency is limited to $22,050$ Hz based on the sampling theorem. The depth map is estimated by using a depth estimation network from the obtained spectrograms. For the depth estimation network architecture, we use exactly the same architecture as EchoNet used in a recent echo-based depth estimation method~\cite{parida2021beyond}, which is an encoder-decoder type CNN consisting of three convolution layers and seven deconvolution layers. We train it by Adam for $300$ epochs with the batch size of 8 and the learning rate of $0.0001$.
The CNN is trained to recover the ground truth depth map from the spectrograms by minimizing the RMSE between the estimated and the ground truth depth maps.

\begin{figure}[t]
    \centering
    \includegraphics[width=0.9\linewidth]{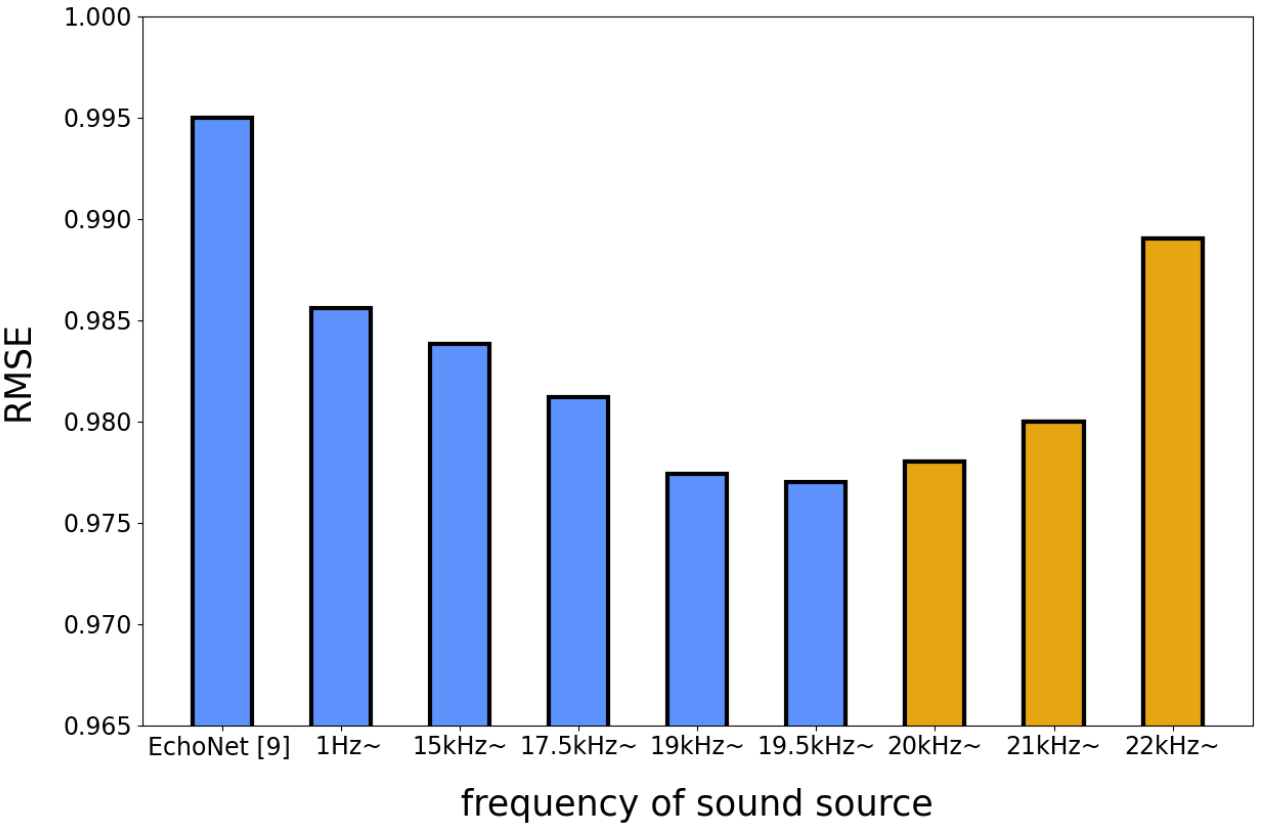}
    \vspace{-12pt}
    \caption{{\bf Results of Preliminary Experiments.} RMSE values of all the frequency setups (lower is better). The blue and orange bars indicate the results using audible and ultrasonic sound sources, respectively.}
    \vspace{-12pt}
    \label{fig:gosa}
\end{figure}

\subsection{Results}
We evaluate the depth estimation accuracy in terms of RMSE between the estimated and ground truth depth maps. We basically follow \cite{parida2021beyond} on the training protocols.
We report the average of five runs with different random seeds.
To evaluate the estimation accuracy with echoes in different frequency bands, we use the following eight cutoff settings of the high-pass filter: $1$ Hz, $15,000$ Hz, $17,500$ Hz, $19,000$ Hz, $19,500$ Hz, $20,000$ Hz, $21,000$ Hz, $22,000$ Hz. $20,000$ Hz and above are considered ultrasonic, so three out of the eight settings of the bands correspond to the ultrasonic cases.

The results are shown in Fig.~\ref{fig:gosa}. Up to $19,500$ Hz, the accuracy tends to be improved (i.e., RMSE decreases) as the frequency band is limited. 
This may be due to the dominance of the high-frequency band, which has high measurement accuracy in theory. 
However, as the frequency band is further restricted above $20,000$ Hz, the accuracy is observed to decrease. 
The reason may be due to a decrease in power in the ultrasonic band due to the effect of attenuation and a decrease in the amount of information due to the limitation of the frequency band.
To conclude, we observed that the depth estimation accuracy decreased with the ultrasonic band only, and improved with a slightly lower frequency band included.

\begin{figure}[t]
    \centering
    \includegraphics[width=0.9\linewidth]{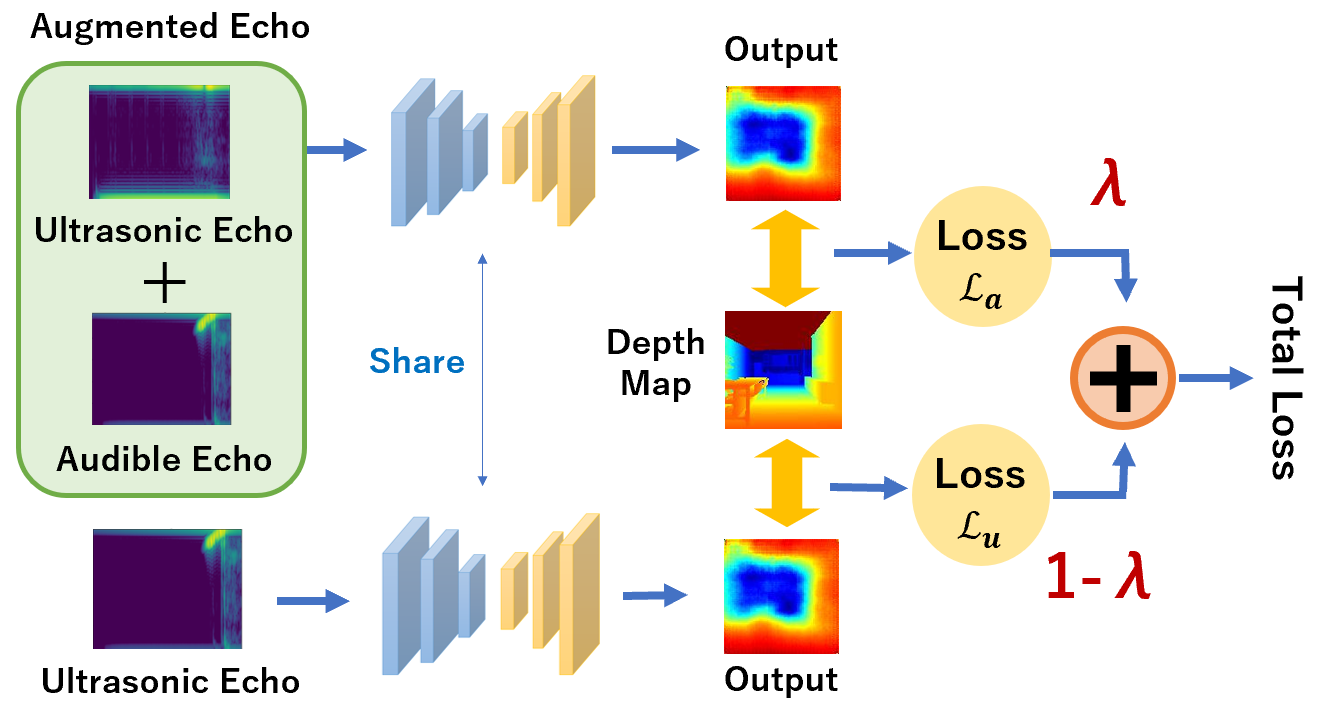}
    \vspace*{-0.4cm}
    \caption{{\bf Our Method.} Generate an augmented echo by fusing an ultrasonic echo and an audible echo with a lower frequency band. 
    Learning is performed to minimize the weighted sum of the two losses evaluated for the two depth maps estimated using ultrasonic and augmented echoes, respectively. 
    The weight $\lambda$ is scheduled as the learning proceeds.}
    \label{fig:teian}
    \vspace*{-0.3cm}
\end{figure}

\section{Method} \label{sec:approach}
Based on the results of the preliminary experiments reported above, we propose a novel deep learning method to improve ultrasonic echo-based depth estimation.
The overview of the proposed method is illustrated in Fig.~\ref{fig:teian}. 
The key idea is to use audible echoes obtained from lower-frequency audible sounds only during training, with the aim of obtaining a depth estimation network robust to missing lower-frequency bands.

\subsection{Auxiliary Echo Generation}
The core of the proposed method is to generate an ``augmented echo" synthesized by combining the spectrograms of the ultrasonic echo obtained from an ultrasonic source and of an auxiliary lower-frequency echo obtained from an audible source. 
The combination is performed in the Mixup data augmentation manner~\cite{zhang2018mixup} as follows.
\begin{eqnarray}
X_a &=& \alpha X_{u} + (1 - \alpha) X_{l}, \\
Y_a &=& \alpha Y_{u} + (1 - \alpha) Y_{l}.
\end{eqnarray}
$X_u, X_l$ are the two spectrograms to be mixed; in our method, $X_u$ and $X_l$ are the spectrograms of the ultrasonic echo and auxiliary lower-frequency echo, respectively.
$Y_u, Y_l$ are the corresponding ground truth depth maps. 
$X_a, Y_a$ are the synthesized spectrogram of the augmented echo and the ground truth depth map, respectively. 
$\alpha$ is the mixing ratio drawn from the uniform distribution on $[0, 1]$.
Note that our method always synthesizes two echoes observed at the same location and orientation, so the ground truth depth maps to be mixed are exactly the same, i.e.., $Y_a = Y_u = Y_l$. Hence, it is not necessary to explicitly mix the depth maps.
Since there is a concern that mixing two spectrograms with significantly different frequency bands may not provide effective augmented echoes for training, the proposed method limits the bandwidth difference between $X_u$ and $X_l$ to 1,000 Hz or less.

\begin{figure}[t]
    \centering
    \includegraphics[width=0.9\linewidth]{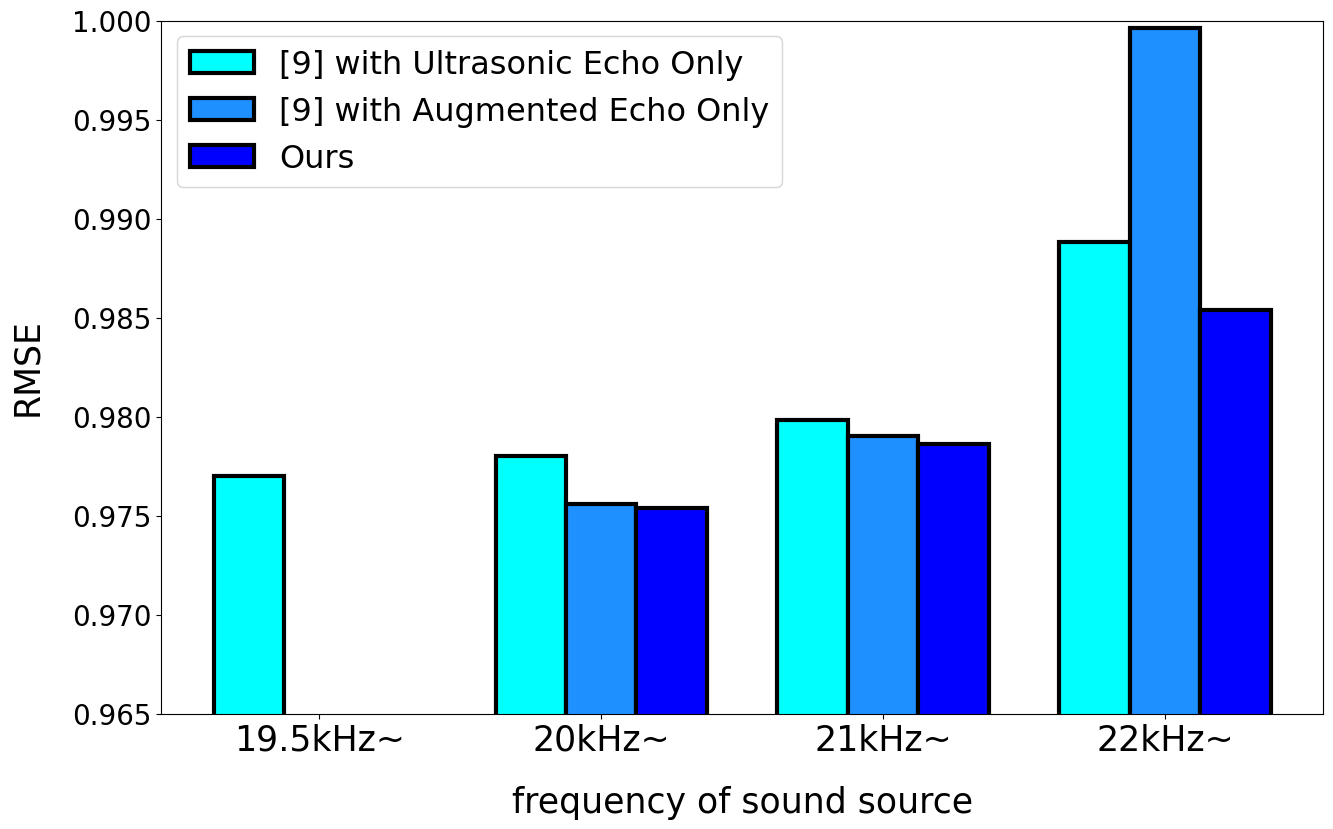}
    \vspace*{-0.4cm}
    \caption{{\bf Quantitative Results.} RMSE values of the ultrasonic echo only, augmented echo only, and the proposed method (lower is better).}
    \label{fig:kekka1}
    \vspace*{-0.5cm}
\end{figure}

\subsection{Loss Configuration}
Our depth estimation network is trained with the RMSE loss. However, training the network with only the augmented echoes is not desirable.
The final goal is to improve the depth estimation accuracy when only the ultrasound echo is fed into the network. 
Meanwhile, the synthesized augmented echoes always contain audible band spectra (except for the rare case where $\alpha=1$), which are useful in the early stages of training but not suitable for its later stages.
Based on this idea, the proposed method uses a total loss function defined as a weighted combination of two (sub-)loss functions, one for ultrasonic echoes and the other for augmented echoes, and schedules the weight as the learning proceeds.

Let $\mathcal{L}_{u}(X_u, Y_u)$ be the loss function for ultrasonic echoes and $\mathcal{L}_{a}(X_a, Y_a)$ be that for augmented echoes. Our total loss function $\mathcal{L}(X_u, X_a, Y_a)$ is defined as:
\begin{equation}
\mathcal{L}(X_u, X_a, Y_a) = \lambda \mathcal{L}_{a}(X_a, Y_a) + (1-\lambda) \mathcal{L}_{u}(X_u, Y_a),
\end{equation}
where $\lambda$ is a hyperparameter to control the balance between the two loss functions. 
This configuration allows us to always evaluate the loss when only ultrasonic echoes are used. $\lambda$ is scheduled as the learning progresses by changing from $\lambda=1$ to $0$ in a linear scheduling manner.

\section{EXPERIMENTS} \label{sec:experiments}

We evaluate the depth estimation accuracy of our method for three frequency settings of the ultrasound band, i.e., when the cutoff frequency of the high-pass filter is set to $20,000$ Hz, $21,000$ Hz, and $22,000$ Hz. The cutoff frequencies of the corresponding auxiliary lower-frequency echoes are set to $19,500$ Hz, $20,000$ Hz, and $21,000$ Hz.
For comparison, we evaluate two baselines: the method in \cite{parida2021beyond} applied to ultrasonic and augmented echoes, respectively.
Other experimental conditions are the same as those used in our preliminary experiments (see Sec.~\ref{sec:prelimininary}).

\begin{figure}[t]
    \centering\includegraphics[width=0.7\linewidth]{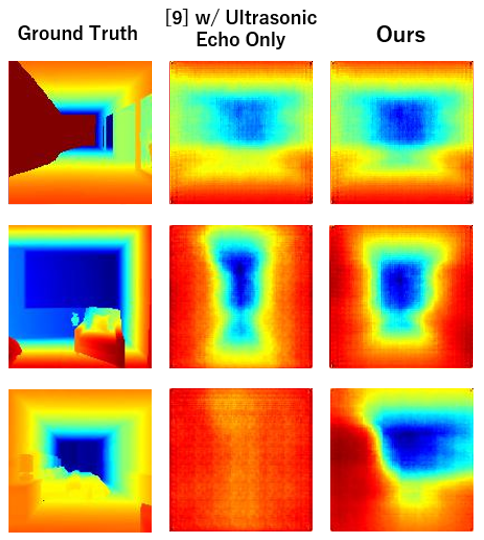}
    \vspace*{-0.5cm}
    \caption{{\bf Qualitative Results.} 
    From left to right, the ground truth depth maps, the results obtained by applying \cite{parida2021beyond} to ultrasonic echoes, and the results by our method.}
    \label{fig:depth}
    \vspace*{-0.5cm}
\end{figure}

\subsection{Quantitative Results}
The quantitative results are shown in Fig.~\ref{fig:kekka1}.
Our method achieves the best accuracy for all the settings. 
First, \cite{parida2021beyond} with augmented echoes surpasses that with ultrasonic echo, which demonstrates the effectiveness of using augmented echoes during learning.
Second, Ours is better than \cite{parida2021beyond} with augmented echoes. This is because low-frequency components are always mixed into the training data in \cite{parida2021beyond} with augmented echoes, resulting in overfitting to low-frequency bands that are not actually used for estimation.
These results verify the effectiveness of the proposed method.

\subsection{Qualitative Results}
A few examples of depth maps estimated by the proposed method are shown in Fig.~\ref{fig:depth}. The estimated depth maps by ours are closer to the ground truth depth maps than those estimated by \cite{parida2021beyond} with ultrasonic echoes. This result further emphasizes the superiority of the proposed method.

\section{CONCLUSION} \label{sec:conclusion}
We explored ultrasonic echo-based depth estimation. We proposed a novel method of transferring knowledge of audible sound to ultrasound, based on the solid support of our analysis (Fig.~\ref{fig:gosa}). As the first paper addressing ultrasound-based depth estimation, we believe that this paper can provide a new direction to the community. Performance evaluation on real datasets based on our proposed method is a future work.

\vspace{5pt}
\noindent {\bf Acknowledgment.} This work was partially supported by JSPS KAKENHI Grant Number 23K11154.





\bibliographystyle{IEEEbib}
\vspace*{-0.3cm}
\bibliography{main.bib}
\vspace*{-0.3cm}


\end{document}